# Dual-resonance nanostructures for colour down-conversion of colloidal quantum emitters

*Son Tung Ha,[1,†,*] Emmanuel Lassalle,[1,†] Xiao Liang,[2] Thi Thu Ha Do,[1] Ian Foo,[1] Sushant Shendre,[2] Emek Goksu Durmusoglu,[2,3] Vytautas Valuckas,[1] Sourav Adhikary,[1] Ramon Paniagua-Dominguez,[1] Hilmi Volkan Demir,[2,3,4*] and Arseniy Kuznetsov[1*]*

[1] Institute of Materials Research and Engineering, Agency for Science Technology and Research (A*STAR), 2 Fusionopolis Way, Singapore 138634

[2] LUMINOUS! Center of Excellence for Semiconductor Lighting and Displays, The Photonics Institute, School of Electrical and Electronic Engineering, Nanyang Technological University, 639798, Singapore

[3] Division of Physics and Applied Physics, School of Physical and Mathematical Sciences, Nanyang Technological University, 21 Nanyang Link, Singapore 637371

[4] UNAM—Institute of Materials Science and Nanotechnology, The National Nanotechnology Research Center, Department of Electrical and Electronics Engineering, Department of Physics, Bilkent University, Bilkent, Ankara, 06800, Turkey

## ABSTRACT

Linear colour conversion is a process where an emitter absorbs a photon and then emits another photon with either higher or lower energy, corresponding to up- or down conversion, respectively. In this regard, the presence of a volumetric cavity plays a crucial role in enhancing absorption and photoluminescence (PL), as it allows for large volumes of interaction between the exciting photons and the emissive materials, maximising the colour conversion efficiency. Here, we present a dual resonance nanostructure made of a titanium dioxide (TiO$_2$) subwavelength grating to enhance the colour down-conversion efficiency of green light at ~530 nm emitted by gradient alloyed Cd$_x$Zn$_{1-x}$Se$_y$S$_{1-y}$ colloidal quantum dots (QDs) when excited with a blue light at ~460 nm. A large mode volume can be created within the QD layer by the hybridisation of the grating resonances and waveguide modes. This allows increasing mode overlap between the resonances and the QDs, resulting in large absorption and tailored emission enhancements. Particularly, we achieved polarized light emission with maximum photoluminescence enhancement of ~140 times at a specific angular direction, and a total enhancement of ~34 times within 0.55 numerical aperture (NA) of the collecting objective. The enhancement encompasses absorption enhancement, Purcell enhancement and directionality enhancement (i.e., outcoupling). We achieved total absorption of 35% for green QDs with a remarkably thin colour conversion layer of ~ 400 nm (inclusive of the TiO$_2$ layer). This work provides a guideline for designing large-volume cavities for practical application in absorption/fluorescence enhancement, such as down colour conversion in microLED displays, detectors or photovoltaics.

**INTRODUCTION**

Augmented reality (AR) devices such as smart glasses are called to be the "next revolution" in consumer electronics. Tech giants such as META, Apple, Google, and Xiaomi have heavily invested in their versions of these devices with the hope of completely changing the way people interact with the virtual world and bringing a true metaverse experience to the users. Compared to more matured display technologies such as liquid crystal display (LCD) and organic light emitting diode (OLED), microLED based on III-V semiconductors are the emerging technology believed to be the best solution in this arena, thanks to its superior performance in brightness, stability, contrast, and colour gamut.[1,2] Beyond AR, microLED can be a game changer in display technology for many applications such as wearable devices, virtual reality (VR), micro projectors, etc. However, this technology is still at its infancy, and there are many challenges to be addressed before we can see these devices on the market.[3] Today, the main obstacle is to accurately assemble the millions of micro LEDs necessary for ultra-high definition displays on the backplane with a near-zero tolerance for error. This problem becomes dramatically more complicated when it comes to a full-colour display where three separate colour channels (i.e., red, green, and blue - RGB) are needed to be precisely placed. This assembling issue, together with the intrinsic challenges in green and red microLEDs, such as low efficiency of green microLEDs, high surface recombination in red microLEDs, and different driving voltages compared to blue microLED, makes microLED display technology hardly achievable for consumer products. To address these challenges, both the research and industry communities are now turning to an alternative approach based on a monolithic blue microLED backplane and a colour down-conversion layer for green and red using colloidal QDs[4–8] (Figure 1a).

The simplest way to integrate green or red QDs onto a microLED display is to place them directly on top of a blue microLED pixel made of gallium nitride (GaN), where a green or red pixel is supposed to be. The principle is that the QDs will absorb the blue light from the microLED and then emit either green or red colour. However, there are challenges associated with the colour conversion approach, such as the leaking of blue light when it is not fully absorbed in the QD layer, affecting the colour purity, the cross-talk between colour pixels, and the patterning of the QD layer to large aspect ratios.[9–16] The latter is necessary due to the relatively small absorption cross-section of QDs, especially with the green ones which require a thick layer (i.e., tens of microns) to completely absorb the blue light from the sourced microLEDs. Patterning becomes increasingly challenging when the pixel size is reduced to a few microns, as required for high-resolution display applications.[1,2,15] In this regard, a resonant structures can help to solve the above-mentioned issues by enhancing the absorption of the QD layer and managing its emission directionality.[17–20] There have been numerous studies in the nanophotonics community to use resonant nanostructures to enhance the fluorescence of emitters based on the Purcell effect.[21–25] However, these previous studies have one thing in common, a relatively small mode volume of the resonant structures. This means that only a small amount of QDs within the "hotspots" can couple to the mode and thus be enhanced. These structures cannot be applied in the colour down-conversion in microLED displays where a large amount of QDs is needed to completely absorb the blue light and the emitters have a good quantum yield, to begin with. Another commonly used resonant approach is based on the Fabry-Perot (FP) cavities. By simply sandwiching the QD layer in between two Diffraction Bragg Reflector (DBR) mirrors which circulate the blue light within the layer, absorption of QDs can be substantially enhanced, thus improving the colour conversion efficiency.[10,13,26] However, to fabricate DBR top mirrors, deposition techniques such as sputtering or thermal evaporation are required. The process usually leads to damaging the QD layer beneath the mirror.[10] In addition, these DBRs require a homogeneous QDs thickness across the pixel and an optically smooth surface, which are typically difficult to achieve by a solution-based patterning method. A few other approaches have been studied to enhance colour conversion efficiency, such as the incorporation of QDs into InGaN nanorod LEDs;[27] mesoporous QD layer;[28] and incorporating high refractive index nano-scatterers (e.g., $TiO_2$).[8,29–31] However, these approaches have drawbacks such as low enhancement efficiency, difficulty incorporating QDs into the structures, and requiring modification of microLED structure (e.g., in InGaN nanorods). Thus, there is a need to develop a resonant structure with both large mode volume and relaxed conditions for QDs integration in the colour downconversion technology.

In this work, we propose a resonant nanostructure based on the hybridisation of a waveguide mode within the QDs layer and a resonant "Bloch mode" originating from the high-index contrast grating made of $TiO_2$ stripes, as shown in Figure 1b. The resonant structure is designed to have dual resonances at the blue wavelength (emission peak of microLEDs) and green wavelength (emission peak of green QDs), which can enhance both in-coupling (i.e., absorption) and outcoupling (i.e., emission) of polarized light. Most importantly, it has a large mode volume that results in the photoluminescence enhancement of a large number of QDs within the structure. In this work, we chose green QDs (i.e., emission peak centered at ~530 nm) as a case study because of their lower absorption cross-section at the blue wavelength (i.e., 460 nm) compared to red QDs and, thus, a higher challenge to achieve the full color conversion. However, this design concept can be applied to the red QDs as well and is suitable for full-colour display applications.

**RESULTS AND DISCUSSION**

**Design principle, fabrication and integration of QDs into the resonant nanostructure**

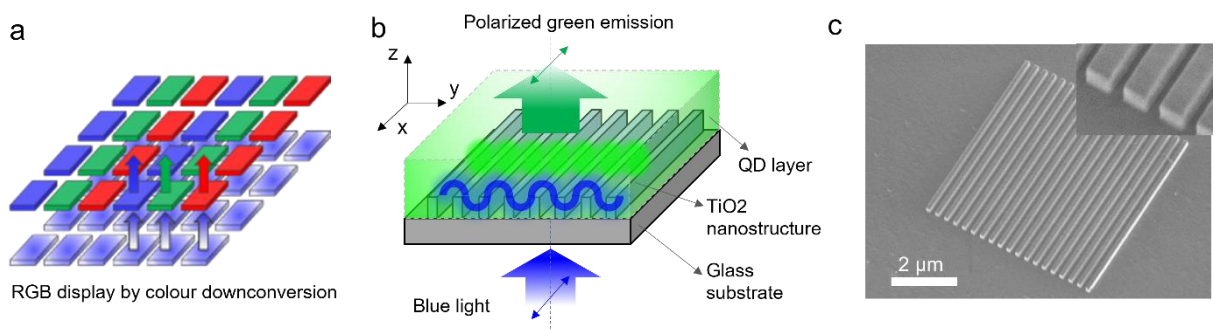

**Figure 1.** (a) Concept of RGB full-colour display by down-colour conversion using a blue microLED monolithic backplane. (b) Dual-resonance nanoantenna-assisted down-colour conversion concept used in this work. (c) SEM images of the fabricated $TiO_2$ nanostructure. The inset shows a "zoomed in" view of an edge of the nanostructure to highlight the quality of the side walls.

We chose TiO$_2$ as the grating stripe material because of its relatively high refractive index and negligible optical loss in the visible.[32] It is well known that such high-index contrast grating may support "Bloch mode resonances" (also called "grating resonances") with strong angular response and polarisation effects,[33–36] which we will exploit here to mediate the absorption and emission of light by QDs. While the strong angular response offered by such gratings can be an advantage for emission, to make it more directional into a narrow angular cone and enhance the collection efficiency, efficient absorption often needs a rather wide angular response, especially if the source has a nearly-isotropic emission (e.g. Lambertian light sources). This can be engineered by carefully controlling another parameter in this design, which is the thickness of the QD layer, in order to support a single waveguide/slab mode band near the wavelength of absorption. In this case, the strong coupling between the slab mode (which is excited by the diffraction from the grating) and the grating/lattice resonance can lead to an avoided crossing resulting in a "flattening" of the band over a wide angular range.[37] In addition to this, the hybridisation of the modes leads to a higher mode overlap with the gain medium leading to a higher absorption, as will be shown later.

We designed the TiO$_2$ grating to match two grating resonances (for normal incidence) with the blue ($\lambda = 460$ nm) and green ($\lambda = 530$ nm) light wavelengths, which those of interest for absorption and emission, respectively. To do so, the resulting grating has a period $\Lambda = 270$ nm, and the stripes have a height $H = 200$ nm and a width $W = 190$ nm. The total thickness of the QD layer is designed to be $T = 400$ nm (i.e., 200 nm additional layer on top of the stripes) to support the first-order transverse electric (TE) slab mode, TE$_1$, around 460 nm. For more details regarding the design, see Supp. Info. Section 1, Figure S1-3. The measured optical constants of TiO$_2$ and the QDs are given in the Supp. Info. Section 2, Figures S4-5. We then fabricate the TiO$_2$ grating structures with different array sizes (i.e., 5×5, 10×10, 20×20, and 50×50 μm$^2$) using e-beam lithography and dry etching processes. The detailed nanofabrication parameters can be found in the Methods section. Figure 1c shows SEM images of the fabricated 5×5 µm$^2$ grating structure. QDs are then spin-coated onto the fabricated sample. The concentration of QDs and spin-coating speed are controlled to yield ~ 400 nm total thickness of the QDs/TiO$_2$ grating structure as measured by a profilometer (Supp. Info. Section 2, Figure S6).

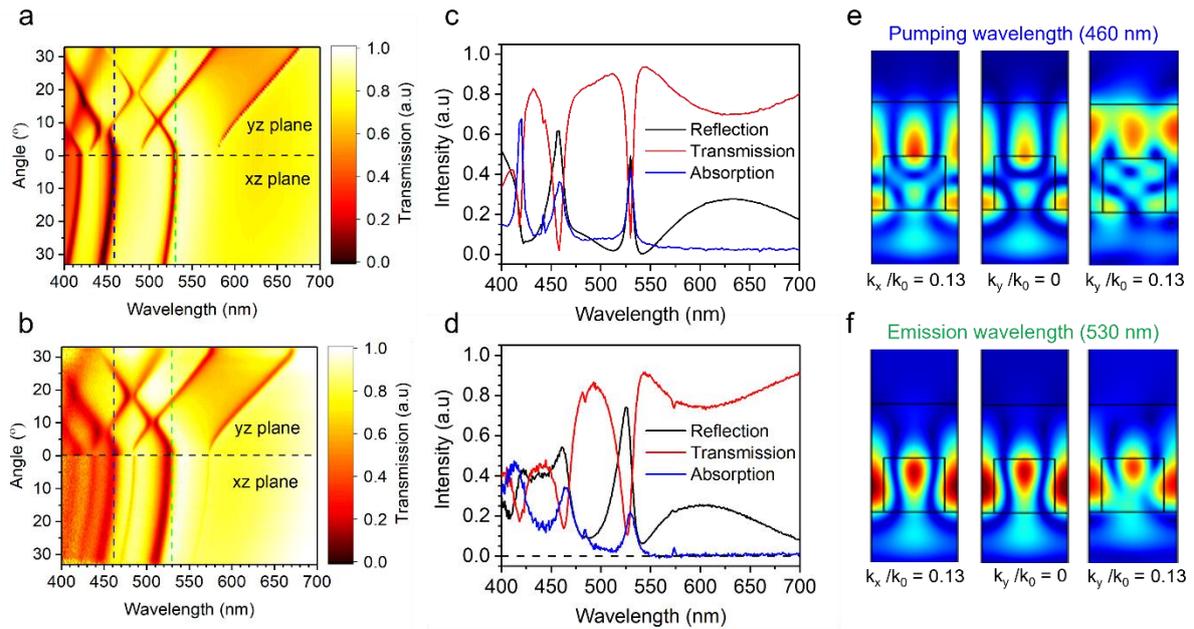

**Figure 2.** (a-b) Simulated (a) and measured (b) angle-resolved transmission spectra, for an incident light coming from the glass substrate side with linear polarization along the stripe direction (i.e., along x axis), in two different planes of incidence: xz plane (bottom halfs of the panels) and yz planes (top halfs of the panels). (c-d) Simulated (c) and measured (d) absorption spectrum for normally incident excitation (blue) calculated from transmission (red) and reflection (black) spectra. (e) Field distribution at pumping wavelength (i.e., 460 nm) at different k values. (f) Field distribution at emission wavelength (i.e., 530 nm) at different k values defined in air as sin ($\theta$) with $\theta$ being the angle of incidence.

The optical properties of the samples are characterised using a customized back focal plane micro-spectrometer (the detailed measurement setup is described in the Methods section and elsewhere).[38] Figures 2 a,b show both the simulated and experimentally measured angle-resolved transmission spectra when the system is excited with incident light polarized along the stripes (i.e., x axis), and in two different planes of incidence: the $yz$-plane (top half of the panel) and the $xz$-plane (bottom half of the panel). The presence of the resonances can be seen as "dips" in the transmission spectra, as shown in Figure 2c,d (in red) which represent a cut of the angle-resolved spectra at normal incidence (0°). In particular, one can clearly see the two resonance bands at $\lambda = 460$ nm and $\lambda = 530$ nm. For the former, the avoiding crossing between the grating resonance and the slab mode is evident at incident angle of ~7.3° in yz plane (i.e., $k_y/k_0 = 0.13$), which is a signature of the strong coupling and hybridisation between the two modes.[34,35] The strong coupling of these two resonance modes leads to a "flat" band over $k_y/k_0 = \pm 0.13$ in the $yz$ plane, which constitutes a total angular range of ~15° in that plane. For completeness, the simulated angle-resolved transmission spectra of the system when excited with incident light polarized perpendicular to the stripe (i.e., y axis) are shown in the Supp. Info. Section 3, Figure S7. There, it can be seen that the mode hybridization does not occur, since the design is made to reach the strong coupling condition between modes polarized along the stripes (x axis), i.e. TE Bloch mode and TE waveguide mode (see Supp. Info. Section 1). Note that the transmission measured in the $xz$ plane does not present a strong angular dependence since this plane has no periodicity and, therefore, no diffraction, making it less sensitive to the angles of incidence (and similarly for the other polarization, as seen in Supp. Info. Section 3, Figure S7). To evidence the importance to finely tune the grating parameters to achieve the strong coupling between different resonances, we show in Supp. Info. Section 3, Figure S8 an experimental comparison between grating structures having different stripe widths, and how it affects the avoided crossing behavior (i.e., strong coupling).

We then characterise the absorption spectrum of QDs inside the grating structure, as shown in Figure 2c,d (in blue). The absorption spectrum is calculated as A( $\lambda$ ) = 1 – T($\lambda$) – R($\lambda$), where T and R are the transmission and reflection (red and black curves in Figure 2c,d, respectively).

The measured angle-resolved reflection and transmission spectra used to calculate the absorption can be found in Supp. Info. Section 3, Figure S9. It can be seen that absorption of 35% at 460 nm excitation is achieved for normal incidence, in excellent agreement with the absorption of 36% obtained from simulations (see Supp. Info. Section 4, Figure S10). One of the possible reasons for the moderate absorption is the high reflection (i.e., ~ 55%) at the resonance wavelength of 460 nm. It can also be seen that the reabsorption of QD fluorescence may occur at 530 nm due to absorption enhancement caused by the grating resonance at that wavelength. The details about the numerical simulations can be found in the Methods section.

Further understanding can be gained by looking at the mode profiles (norm of the electric field) at different points in the resonance bands, as shown in Figure 2e,f. While there is already a good overlapping at $\lambda = 460$ nm between the grating resonance and the QD medium at $k_y/k_0 = 0$ (central panel in Figure 2e, and also in the $xz$ plane at $k_x/k_0 = 0.13$, left panel in Figure 2e), which leads to the absorption of 35% observed at normal incidence, the grating resonance clearly gets the nature of a waveguide mode at $k_y/k_0 = 0.13$ (right panel in Figure 2e), which is the result of the strong coupling and mode hybridisation. This leads to an even better spatial overlapping between the resonance mode and the QDs layer, as can be seen in the mode profile. Calculations of the absorption of the QDs at this angle $k_y/k_0 = 0.13$ (7.3°), shown in the Supp. Info. Section 4, Figure S10 reveal that the hybridisation of the resonant Bloch mode with the slab mode leads to a maximum of the total absorption around ~67% at $\lambda \sim 460$ nm (of which 93% comes from absorption of light in the QDs), in addition to the "flattening" of the band described previously. To achieve such a high value of absorption with a bare QD film, one would need an estimated QD film thickness of ~5.8 μm (see Supp. Info. Section S4).

Regarding the band at $\lambda$ ~530 nm, exploited to enhance the emission by the QDs, we also show the corresponding mode profile in Figure 2f. One can see that the electric field still extends outside of the TiO$_2$ stripes, even if the overlap with the gain medium is less than that of the resonance at 460 nm. Modal analysis of the electric and magnetic field components reveals that this resonance at $\lambda = 530$ nm is essentially an in-plane electric dipole mode in nature (along the stripe direction, i.e. along the $x$-axis). We also note that for the other polarization, there is no resonance at $\lambda = 530$ nm (see Supp. Info. Section 3, Figure S7). Therefore, one can expect an enhancement of linearly polarized light emission along x-axis at the QD emission wavelength, as we will see hereafter.

**Colour down conversion performance**

We characterise the colour conversion performance of the fabricated structures by first using a wide-field fluorescence imaging technique. The whole sample was is illuminated by a mercury lamp equipped with a 470 nm short-pass filter through a 10× objective (0.3 NA). The signal is then passed through a 500 nm long-pass filter to collect only the fluorescence from the QDs. Figure 3a shows the wide-field PL images of the fabricated sample. The green pixels correspond to the area where the grating structures are present, while the "dark" area is the bare quantum dot film. It can be seen that the nanoantennas have significantly enhanced the PL of QDs down to pixel size as small as 5×5 µm$^2$. To investigate the effect of the excitation wavelength on the PL enhancement, a tunable unpolarized laser (SuperK, NKT Photonics) is used to excite the QDs at normal incidence. The laser is set to have a fixed bandwidth of 10 nm, and the central wavelength is varied from 450 nm to 470 nm. In all cases, a fixed power of ~500 µW and a laser spot size of 30 µm are used to excite the QDs. The total PL signals are collected using a 50× objective (0.55 NA) with a linear analyser along the stripe direction (i.e., along x axis). The total PL enhancement is calculated by comparing the PL of QDs in the 50×50 µm$^2$ nanoantenna pixel with that of the bare QD film for each excitation condition, as shown in Figure 3b. As expected, the maximum total PL enhancement of ~34 is achieved when the excitation wavelength is set at 460 nm, corresponding to the designed resonance wavelength. By comparing the total PL enhancement when pumping at resonance wavelength (i.e., 460 nm) and non-resonance wavelength (i.e., 470 nm), we can estimate the absorption enhancement of ~3.5 times. The rest of the enhancement is attributed to the Purcell effect and outcoupling enhancement of the emission at the second resonance (i.e., 530 nm), which will be further demonstrated below. Similar experiments are done for different pixel sizes with adjusted laser spot sizes for smaller pixels, and the results are shown in Figure 3c. It can be seen that as the pixel size reduces from 50×50 µm² to 5×5 µm², the total PL enhancement drops from ~34 to about 15. This can be explained by the fact that the resonances are designed based collective modes, which strongly depends on the array size. Nevertheless, a reasonable enhancement can be achieved with a pixel size as small as 5×5 µm², which is enough for most VR and AR applications.

Regarding the emission, Figures 3d,e show the BFP images of the QD emission in a bare film and in the 50-µm pixel, respectively, for a linear polarization along the x-axis. It can be seen that the emission of QDs is concentrated in three lobes in the radiation pattern (Figure 3e). This observation agrees well with the simulated emission pattern of QD in the grating structure, as shown in Figure 3f. The narrower emission band in the simulation compared to the experiment is because the calculation is for a single wavelength at 530 nm rather than the whole emission spectrum. In the radiation pattern, the central lobe is coming from the coupling to the grating resonance at normal incidence which is identified as an electric dipole resonance along the x-axis, as discussed above in Figures 2a,b, while the two side lobes are coming from the coupling to another grating resonance that can be seen at around 16.7° in the $yz$ plane (i.e., $k_y/k_0 = 0.29$ and $k_x/k_0 = 0$) in Figures 2a,b. The simulation results for full radiation patterns of both linear polarizations along the x and y axes (see Methods) can be found in Supp. Info. Figure

S11. As expected, there is no central lobe for linearly polarized light along y-axis, since there is no resonance at normal incidence, while two side lobes can be seen, corresponding to the coupling to another grating resonance that can be observed at around 21.7° in the $yz$ plane (i.e., $k_y/k_0 = 0.37$ and $k_x/k_0 = 0$) in Supp. Info. Section 3, Figure S7. Setting apart the side lobes, the nanoantenna therefore enhances linearly polarized light emission along the x-axis.

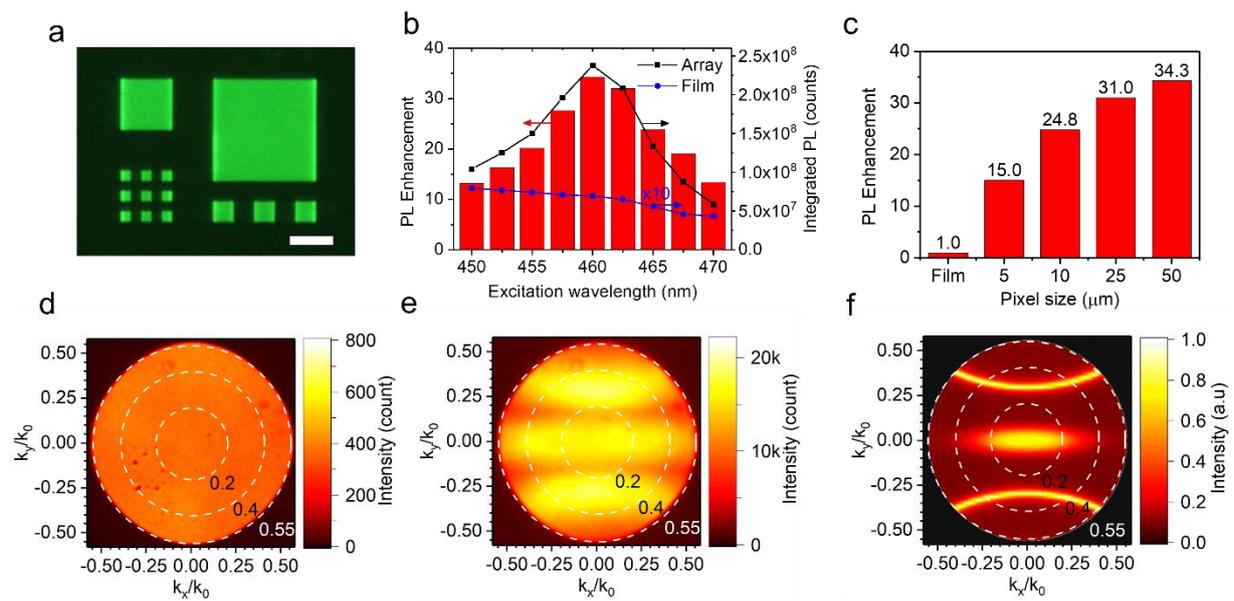

**Figure 3.** (a) Fluorescence image of nanoantenna enhanced colour conversion with different pixel (array) sizes. The scale bar represents 20 µm. (b) Experimental PL enhancement collected with an x-polarized analyzer for different excitation wavelengths for the 50-µm array (red bars), recorded for an unpolarized and normally incident laser excitation. The integrated PL for the nanoantenna array (black dots) and the bare QD film (blue dots) are also shown, for a fixed excitation power and same laser spot size. (c) Experimental PL enhancement collected with an x-polarized analyzer for different pixel sizes when pumping at 460 nm, with all other excitation conditions unchanged. (d-e) Experimental PL BFP images of QDs in a planar film and in a 50 µm pixel, respectively, collected with an x-polarized analyzer. (d) Simulated PL BFP image of the light emission of QDs inside the nanoantennas at 530 nm wavelength linearly polarized along the stripe direction (i.e., x-axis).

To further quantify the outcoupling enhancement of QD emission in the nanoantenna sample, angle-resolved spectroscopy measurements are performed. Figures 4a,b show the spectrally-resolved PL emission in yz plane of the QDs in a bare film and in the nanoantenna structure, respectively. It can be seen again that the PL of QDs couples to two different resonances of the nanoantennas. Figure 4c shows the emission spectra of QDs coupled to these two modes in comparison to that of the bare film of QDs. It can be seen that a maximum PL enhancement of 141 times at 525 nm for the emission angle of ~17° in the yz plane (i.e., $k_y/k_0 = 0.29$ and $k_x/k_0 = 0$) is achieved. PL lifetime measurements are also done at this particular wavelength (i.e., 525 nm), as shown in Figure 4d. It can be seen that the PL lifetime of QDs reduces from 9.8 ns for the bare film to 5.8 ns for the grating structure, indicating a lifetime reduction (i.e., Purcell enhancement).

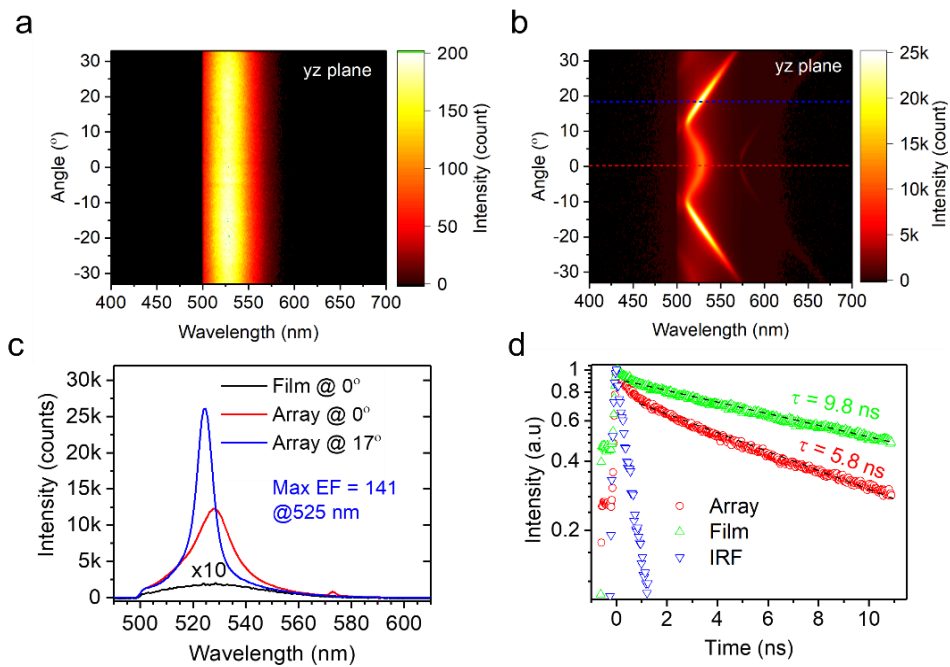

**Figure 4.** (a-b) Experimental angle-resolved PL spectra in yz plane of QDs in a planar film and the grating structure collected with an x-polarized analyzer, respectively. (c) Experimental PL spectra of QDs in the planar film (black line) and the array structure at different emission angles in the *yz* plane (red and blue lines), as shown in b (by horizontal red and blue dashed lines). (d)

Experimental PL lifetimes of QDs in a planar film (green) and the array structure (red). The instrument response function (IRF) is also shown (blue).

## CONCLUSIONS

In conclusion, we present a concept of dual resonance nanostructure with a large mode volume to enhance both the absorption and the light polarized emission of quantum emitters applicable for colour down conversion in display technology. The hybridisation of a grating resonance in the high index contrast grating and a waveguide mode within the QD layer is responsible for better spatial mode overlapping and an angularly "flat-band" resonance at the excitation wavelength over an angular range of 15°. Thus, we report experimentally an absorption of 35% at 460 nm for green QDs for an excitation at normal incidence, which is remarkably high for such a thin colour conversion film of ~400 nm (inclusive of the $TiO_2$ grating layer), while in simulations the total absorption ranges from 36% (at normal incidence) to up to 67% within the angular range of 15°. Moreover, we obtain a highly linear polarized light emission (along the stripes) without the need of filtering, with a total PL enhancement of 34.3 times within the field of view of ± 33° and a maximum enhancement of 141 times at 525 nm wavelength and viewing angle of 17° (i.e., $k_y/k_0 = 0.29$) in comparison to a bare QD film. Our work suggests that high-index dielectric nanoantennas can be used to create efficient resonant structures suitable for colour conversion in practical applications such as microLED displays and sensitiser-assisted detectors.

## METHODS

### Quantum dot synthesis and film preparation

***Chemicals:*** Cadmium oxide (CdO) (99.9%), zinc acetate [$Zn(Ac)_2$] (99.99%), 1-octadecene (ODE), selenium (Se) (99.99% trace metals basis), sulfur (S) (99.98% trace metals basis), oleic acid (OA) (90%), trioctylphosphine (TOP)(90%), octane (98%), hexane, and ethanol (EtOH) were purchased from Sigma Aldrich.

***Synthesis of gradient alloyed $Cd_xZn_{1-x}Se_yS_{1-y}$:*** The green QDs were synthesised according to a method reported in the literature with some modifications.[39] Typically, 0.4 mmol of CdO, 4 mmol of $Zn(Ac)_2$, 5 mL of OA and 15 mL of ODE were mixed and heated to 90°C under vacuum in a 100 mL three-neck flask for 30 mins. Later, the mixture was heated to 310 °C under argon protection. At this temperature, 3 mL of TOP with 0.2 mmol of Se and 4 mmol of S were injected into the flask swiftly. After 10 min, the mixture was cooled down to room temperature in air. For purification, the synthesised QDs were precipitated by adding excess EtOH till the solution became turbid, followed by centrifuging at a speed of 6,000 rpm for 5 min. The above cleaning procedures were repeated twice. The purified QDs with a quantum yield of 80% were dispersed in octane with a concentration of 170 mg/ml for later use.

***Spin-coating of gradient alloyed $Cd_xZn_{1-x}Se_yS_{1-y}$ QDs:*** The substrates with $TiO_2$ nanoantenna structures were cleaned by isopropanol, dried, and treated with oxygen plasma for 1 minutes. Before deposition, the as-prepared QD dispersion was firstly centrifugated at 6,000 rpm for 1 min to remove the large aggregates. Then, the patterned substrate was spin-coated with ~50 μl of the above QDs at 2,000 rpm for 1 min. After that, the as-made QD film was dried under vacuum for 15 min. The samples were finally stored in a dry box before characterisations.

**Nanoantenna fabrication**

***Film deposition:*** $2\times2$ $cm^2$ quartz substrates were first cleaned by deionised (DI) water, acetone, and isopropanol in an ultra-sonication bath. Then amorphos $TiO_2$ film with a thickness of ~200 nm was deposited by ion-assisted sputtering (Oxford Optofab3000). Subsequently, a 30 nm-thick film of chromium (Cr) was deposited by electron-beam (e-beam) evaporation (Angstrom EvoVac).

***E-beam lithography (EBL):*** For nanostructure patterning, a negative electron-beam resist, hydrogen silsesquioxane (HSQ, Dow Corning XR-1541-006) was spin cast on the quartz/$TiO_2$/Cr substrate at 5000 rpm for 60 seconds, followed by baking at 120 °C for 2 minutes and subsequently at 180 °C for another 2 minutes. This results in a HSQ film thickness of ~150 nm. The sample then underwent e-beam lithography (Elionix ELS-7000) with a voltage of 100 keV and a current of 500 pA. Finally, exposed HSQ patterns was developed in a home-made salty developer (1 wt.% NaOH and 4 wt.% NaCl in DI water) for 4 minutes, followed by a generous rinse in DI water.

***Etching of $TiO_2$ nanostructures:*** Inductively coupled plasma reactive ion etching (ICP-RIE, Oxford PlasmaPro 100 Cobra) was used to transfer the EBL patterns to Cr layer using a mixture of $Cl_2$ and $O_2$ gases ($Cl_2$: 20 sccm, $O_2$: 2 sccm, 14 mTorr). Subsequently, using Cr as a hard mask, the pattern was transferred to the $TiO_2$ layer using $CHF_3$ gas (25 sccm, 25 mTorr). Finally, Cr together with any residual HSQ were removed by immersing the sample in liquid chromium etchant (Sigma Aldrich) for 4 minutes. The sample was rinsed in water and IPA and blown dry with nitrogen gas.

**Numerical simulations**

All the numerical simulations were performed using COMSOL Multiphysics. We built a 2D model in the $yz$ plane (because of the invariance of the stripes along the $x$ direction), and simulated only a unit-cell containing one stripe, with the use of periodic boundary conditions in the $y$ direction and of ports in the $z$ direction.

***Transmission/Reflection/Absorption simulations:***

For these simulations (Figs. 2a,c in the main text and Figs. S7 and S10 in Supp. Info.), the incident light was linearly polarized and coming from the glass substrate (refractive index $n_s = 1.46$) with different angles of incidence $\theta$ with respect to the $z$ axis, the plane of incidence being either in the $xz$ plane or in the $yz$ plane, and a sweep over wavelengths was performed. The total reflection $R$ was recorded by the lower port (in glass $n_s = 1.46$) and the total transmission $T$ by the upper port (in air $n_{air} = 1$). In order to compare with the experimental measurements where the angle of incidence was defined in air (i.e., before entering the glass substrate), we converted the angles of incidence in our simulations to: $\arcsin\left[\frac{n_s}{n_{air}}\sin(\theta)\right] \simeq 1.46\,\theta$ (approximation valid for small angles). The total absorption $A$ was obtained by energy conservation as $A = 1 - T - R$. To calculate the different absorption contributions coming from TiO$_2$ or QDs (i.e., the only materials with non-zero extinction coefficient $\kappa$), we integrated the absorption within the TiO$_2$ or QD regions $S_{TiO_2}$ and $S_{QD}$, respectively, according to the formula:

$$Q_i = \frac{1}{C}\iint_{r\in S_i}|E(r)|^2 \text{Im}[\epsilon(r)]dr \quad \text{with} \quad S_i = S_{TiO_2} \text{ or } S_i = S_{QD}$$

using the normalization constant $C$ such that $A = Q_{TiO2} + Q_{QD}$.

***Back focal plane emission and emission enhancement simulations:***

The BFP emission simulations (Fig. 3f in main text and Fig. S11 in Supp. Info.) were performed based on the reciprocity principle.[40,41] Plane waves with two orthogonal linear polarizations (electric field in the $xz$ and $yz$ planes, respectively) were sent from the air side with all different wavevectors in the $xz$ and $yz$ planes, at wavelength $\lambda = 530$ nm, and the electric field power was recorded in the QD region (and averaged over position and orientation). By reciprocity, for a given couple of wavevectors $(k_x, k_y)$, this power corresponds to the power emitted in this same direction by the QD ensemble. This power $P_{QD}$ computed in the nanoantenna case was then normalized by the power $P_{ref}$ emitted by the QDs in the reference case consisting of a film of QDs of thickness 200 nm on top of a glass substrate, and therefore the plotted quantities correspond to the emission enhancement for a given direction and polarization.

**Optical characterisations**

*Back focal plane measurements:* Angle-resolved spectroscopy measurements were performed in a home-built micro-spectrometer setup, consisting of an inverted optical microscope (Nikon Ti-U) coupled to a spectrometer (Andor SR-303i) equipped with an 2-dimensional (2D) electron-multiplying charged-coupled detector (EMCCD, Andor Newton 971, 400 × 1600 pixels). For reflectance measurements, light from a lower halogen lamp of the microscope passing through a linear polarizer was focused onto the sample surface via a 50x microscope objective (Nikon, NA = 0.55). The reflected light was collected with the same objective and passed through a series of lenses, which imaged the back focal plane (BFP) of the objective onto the entrance slit of the spectrometer. The slit width was set to 100 μm to capture a thin slice of the BFP image. A grating with groove density of 150 gr/mm was used to disperse the wavelengths. The 2D-CCD in the spectrometer then gave a reflectance spectrum resolved in both wavelength and angle of reflection. For transmission measurements, light from an upper halogen lamp of the microscope was focused onto the sample via a 50x microscope objective (Nikon, NA = 0.55) and the signal was collected by another 50x microscope objective (Nikon, NA = 0.55), same as was used for the reflection measurements. For PL measurements, light from a supercontinuum light source (superK, NKT photonics) was focused onto the sample via a 50x microscope objective (NA = 0.55), and the PL from the sample was collected through the same objective.

*Characterisation of colloidal QDs:* Photoluminescent spectra of the as-synthesised QD dispersion and spin-coated QD films were measured by a spectra-fluorophotometer (Shimadzu, RF-5301PC) with an excitation wavelength of 380 nm. The absorption spectra of the QD dispersion were recorded by an ultraviolet-visible spectrophotometer (Shimadzu, UV-1800). The quantum yield of the QDs was recorded by an integrating sphere by calculating the ratio between the absolute emission and absorption photons. The transmission electron microscope (TEM) images of the QDs were obtained using a JEOL 2100F TEM at 200 KV.

## ASSOCIATED CONTENT

**Supporting Information** (Design principles, additional sample and optical characterisations, and absorption calculations) is available free of charge.

## AUTHOR INFORMATION

### Corresponding Authors

Son Tung Ha: ha_son_tung@imre.a-star.edu.sg

Hilmi Volkan Demir: hvdemir@ntu.edu.sg

Arseniy Kuznetsov: arseniy_kuznetsov@imre.a-star.edu.sg

### Author Contributions

S.T.H, R.P.D, and A.I.K initiated the idea of the project. R. P-D proposed the conceptual design, I.F did initial design and simulations, and E.L finalized the design and did all other simulations supervised by R.P.D. S.T.H and D.T.T.H fabricated the $TiO_2$ nanostructures and did all back focal plane measurements. V.V and S.A deposited $TiO_2$ films. X.L, S.S, E.G.D, and H.V.D synthesised and characterised QDs, and prepared QD films. S.T.H, H.V.D and A.I.K supervised the project. The manuscript was written through contributions of all authors. All authors have given approval to the final version of the manuscript. †These authors contributed equally.

## ACKNOWLEDGEMENTS

Authors acknowledges the funding support from Singapore MTC-Programmatic Grant No: M21J9b0085. S.T.H also acknowledges the support under the Singapore AME Young Individual Research Grant No. A2084c0177.

# Supplementary Information for

# Dual-resonance nanostructures for colour down-conversion of colloidal quantum emitters

Son Tung Ha,[1,†,*] Emmanuel Lassalle,[1,†] Xiao Liang,[2] Thi Thu Ha Do,[1] Ian Foo,[1] Sushant Shendre,[2] Emek Goksu Durmusoglu,[2,3] Vytautas Valuckas,[1] Sourav Adhikary,[1] Ramon Paniagua-Dominguez,[1] Hilmi Volkan Demir,[2,3,4*] and Arseniy Kuznetsov[1*]

[1] Institute of Materials Research and Engineering, Agency for Science Technology and Research (A*STAR), 2 Fusionopolis Way, Singapore 138634

[2] LUMINOUS! Center of Excellence for Semiconductor Lighting and Displays, The Photonics Institute, School of Electrical and Electronic Engineering, Nanyang Technological University, 639798, Singapore

[3] Division of Physics and Applied Physics, School of Physical and Mathematical Sciences, Nanyang Technological University, 21 Nanyang Link, Singapore 637371

[4] UNAM—Institute of Materials Science and Nanotechnology, The National Nanotechnology Research Center, Department of Electrical and Electronics Engineering, Department of Physics, Bilkent University, Bilkent, Ankara, 06800, Turkey

[†]These authors contributed equally to this work

*Corresponding authors: ha_son_tung@imre.a-star.edu.sg, hvdemir@ntu.edu.sg, and arseniy_kuznetsov@imre.a-star.edu.sg

**Table of contents**

## Section 1. Design principles

### A) Resonant dielectric grating

For the design of the dual-resonance nanostructure for colour down-conversion, we consider a subwavelength periodic dielectric grating made of alternate stripes of high ($n_h \sim 2.54$ for titanium dioxide ($TiO_2$)) and low ($n_l \sim 1.73$ for the quantum dots (QDs) used in this work) refractive indices, as depicted in Figure S1a. Such a high-index contrast grating presents optical resonances, also called grating resonances, in the subwavelength regime $\lambda > \lambda_R = n_s \Lambda$, with $n_s$ being the refractive index of the substrate on which the stripes are sitting ($n_s \sim 1.46$ for glass substrate). Physically, these resonances correspond to the excitation of lateral Bloch modes by the *evanescent* diffraction orders.[1,2]

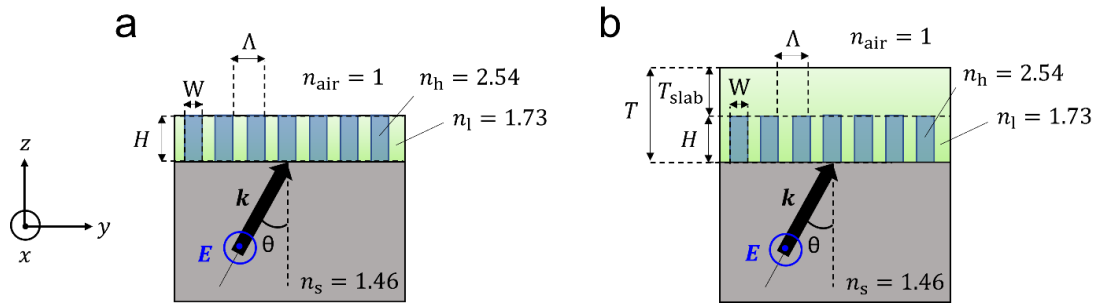

**Figure S1.** (a) Schematics of a periodic grating with period $\Lambda$ in $y$-direction (with infinite number of periods), consisting of stripes (infinitely long in the $x$-direction) of height $H$ and made of alternate high ($n_h = 2.54$) and low ($n_l = 1.73$) refractive index materials, and sitting on a substrate with refractive index $n_s = 1.46$. (b) Same as in a. but with an additional slab of refractive index $n_l = 1.73$ present on top of the grating with a thickness $T_{slab}$. In all the following simulations in this Section 1, the period is set to $\Lambda = 270$ nm, and a linearly polarized light coming from the below substrate is considered, with wavevector $\mathbf{k}$ and electric field $\mathbf{E}$ along $x$-axis. The angle $\theta$ is defined as the angle of incidence with respect to the plane of incidence (e.g. the $yz$ plane in this schematics).

The grating parameters such as the period $\Lambda$ and filling factor $F = W/\Lambda$, ($W$ being the width of the high-index stripes), are chosen to be $\Lambda = 270$ nm and $W = 190$ nm ($F \sim 0.7$) in order to be in the resonant regime for the wavelengths of interest (i.e. for the blue light wavelength $\lambda \sim 460$ nm and green light wavelength $\lambda \sim 530$ nm). This corresponds to the case where lateral Bloch modes are being excited by the *evanescent* first diffraction order. Indeed, the first-order cutoff wavelength $\lambda_c^1$ beyond which one enters the deep subwavelength (non-resonant) regime is $\lambda_c^1 \sim 586.7$ nm for transverse electric (TE) modes (which can be excited by incident light with electric field along the x-axis, as depicted in Figure S1). This value is obtained after solving the transcendental equation (valid for TE modes):[1]

$$n_l \tan\left[\frac{\pi\Lambda}{\lambda}(1-F)n_l\right] = -n_h \tan\left[\frac{\pi\Lambda}{\lambda}F\, n_h\right]$$

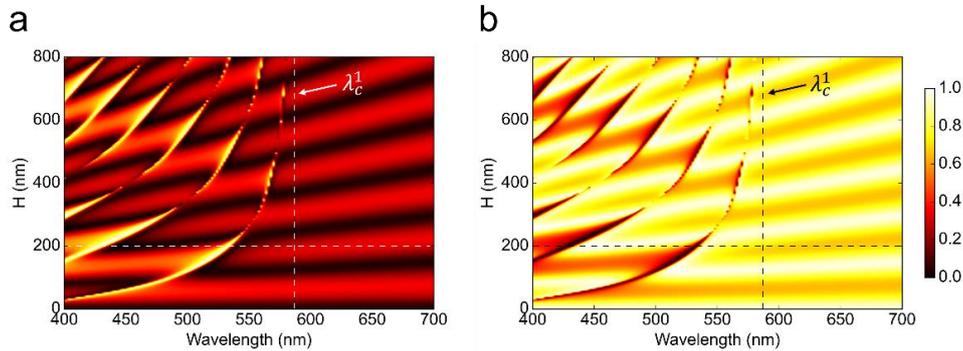

**Figure S2.** (a-b) Simulated reflection and transmission spectra, respectively, as a function of grating height $H$ for normally incident light polarized along $x$-axis (corresponding to the configuration shown in Figure S1a with $\theta = 0°$). The horizontal dashed lines denote the case $H = 200$ nm. The vertical dashed lines denote the first-order cutoff wavelength $\lambda_c^1 \sim 586.7$ nm. In these simulations, constant refractive index values of $n_h = 2.54$ and $n_l = 1.73$ for $TiO_2$ and QDs, respectively, were used (no absorption, i.e. $\kappa = 0$).

Figure S2a (b) show the $0^{th}$-order reflection (transmission) spectrum as a function of the grating height $H$, simulated for $x$-polarized incident light at normal incidence $\theta = 0°$ and coming from the glass substrate. It illustrates the different regimes, the deep subwavelength regime for $\lambda > \lambda_c^1$, where weak Fabry-Perot resonances can be seen (this regime is also called the "homogenization regime" where effective medium theory applies[2]), and resonant regime $\lambda_R < \lambda < \lambda_c^1$ where the strong peaks (dips) in the reflection (transmission) spectrum correspond to the excitation of the lateral Bloch modes by the evanescent first diffraction order, which has an interesting interplay with the Fabry-Perot resonances.

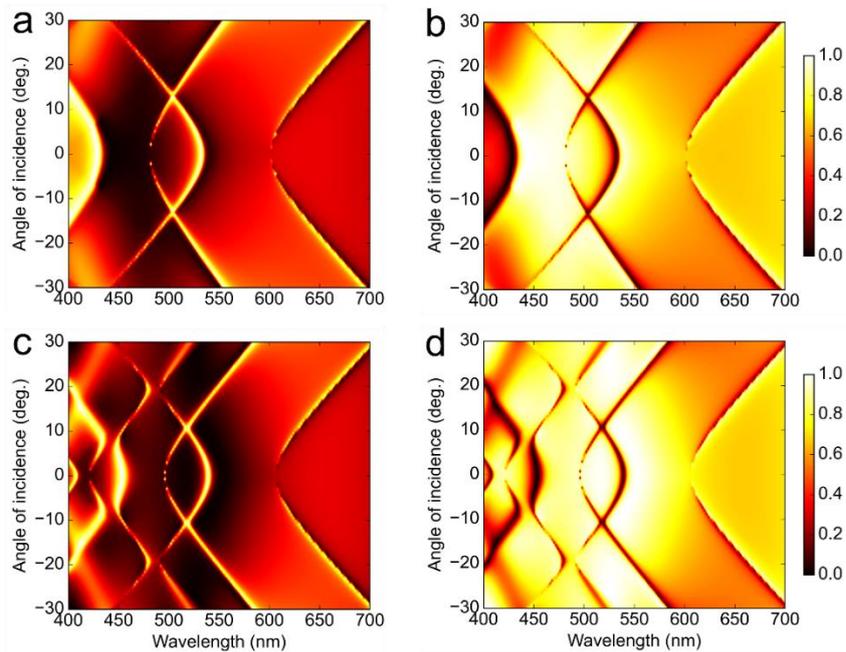

**Figure S3.** (a-b) Simulated angle-resolved reflection and transmission spectra, respectively, for a grating height $H = 200$nm (corresponding to the configuration shown in Figure S1a). (c-d) Angle-resolved reflection and transmission spectra, respectively, for the same configuration but when an additional QDs slab with a thickness $T = 200$nm is present on top of the grating (corresponding to the configuration shown in Figure S1b). In these simulations, constant refractive index values of $n_h = 2.54$ and $n_l = 1.73$ for $TiO_2$ and QDs, respectively, were used (and no absorption, i.e. $\kappa = 0$).

Figure S3a (b) shows the corresponding angle-resolved reflection (transmission) spectrum for a grating structure with $H = 200$ nm, also simulated for $x$-polarized incident light as in Figure

S2, but with varying angles of incidence $\theta$. These "band diagrams" reveal that the lateral Bloch modes are still being excited for off-normal incidences, as well as the presence of two other modes that are not excited at normal incidence due to symmetry constraints (at the Gamma point, these modes form what are known as "symmetry-protected bound states in the continuum", which cannot radiate into the far field).[3]

### B) Slab waveguide

We next interfaced the high-index contrast grating with a slab waveguide made of the low index material, as depicted in Figure S1b. In this subsection, we estimate the minimum slab thickness $T_{\text{slab}}$ necessary to sustain a waveguide mode (the fundamental one) assisted by the grating diffraction. In the case of a 3-layer slab waveguide, made for example of substrate/slab/air layers, the dispersion relation (relation between the frequency $\omega$ and the parallel wavevector $k_\parallel$ of the mode) is obtained by solving a transcendental equation, which in the case of TE waveguide modes (also called slab modes) reads:[4]

$$\tan[\beta T_{\text{slab}}] = \frac{\beta(\gamma + \delta)}{\beta^2 - \gamma\delta}$$

with $\beta = \sqrt{n_{\text{slab}}^2 k_0^2 - k_\parallel^2}$, $\gamma = \sqrt{k_\parallel^2 - n_{\text{sub}}^2 k_0^2}$, $\delta = \sqrt{k_\parallel^2 - n_{\text{air}}^2 k_0^2}$ and $k_0 = \frac{\omega}{c}$.

An approximate solution of this equation can be found by solving $\tan[\beta T_{\text{slab}}] \simeq 0$, which occurs when the argument of the tangent function is equal to multiple of $\pi$, i.e. $\beta T_{\text{slab}} = \alpha\pi$ with $\alpha = 1, 2, 3, \ldots$ an integer labelling the mode order, that is when:

$$\omega_\alpha = \frac{c}{n_{\text{slab}}}\sqrt{\left(\frac{\alpha\pi}{T_{\text{slab}}}\right)^2 + k_\parallel^2}$$

(Note that the case $\alpha = 0$ would represent the case of a plane wave solution in a homogeneous medium of refractive index $n_{\text{slab}}$ and therefore does not describe a waveguide mode.) This equation can be equivalently written in terms of free-space wavelength as:

$$\lambda_\alpha = \frac{2\pi n_{\text{slab}}}{\sqrt{\left(\frac{\alpha \pi}{T_{\text{slab}}}\right)^2 + k_\parallel^2}}$$

When interfacing such slab waveguide with the high-index contrast grating, one can expect resonances (seen e.g. in transmission/reflection spectra…) to occur when the following condition is fulfilled: $k_\parallel = k_\parallel^{\text{inc}} \pm \frac{2\pi m}{\Lambda}$, with $k_\parallel^{\text{inc}}$ the parallel wavevector of the incident light and $m$ the diffraction order ($m = 0, 1, 2…$), which expresses the condition for which the slab mode can be excited by the grating diffraction orders.

One can use the equation above to have an approximate idea of the minimum slab thickness $T_{\text{slab}}^{\text{min}}$ needed to have the fundamental ($\alpha = 1$) slab mode resonance occurring at around $\lambda_{\alpha=1} \sim$ 400 nm, when excited by the first diffraction order $m = 1$ at normal incident light (i.e. $k_\parallel^{\text{inc}} = 0$):

$$T_{\text{slab}}^{\text{min}} = \frac{\pi}{\sqrt{\left(\frac{2\pi n_{\text{slab}}}{\lambda_{\alpha=1}}\right)^2 - \left(\frac{2\pi}{\Lambda}\right)^2}}$$

With the parameters considered here ($\Lambda = 270$ nm, $n_{\text{slab}} = n_l = 1.73$), this gives $T_{\text{slab}}^{\text{min}} \sim$ 200 nm.

Figure S3c(d) shows the angle-resolved reflection (transmission) spectrum simulated for the same grating as in Figure S3a,b but when interfaced with a "slab" of refractive index equal to $n_{\text{slab}} = n_l = 1.73$ and of thickness $T_{\text{slab}} = 200$ nm (situation depicted in Figure S1b). One can see an important modification of the band diagrams compared to Figure S3a,b, with for example a major anti-crossing and band gap opening around $\theta = \pm 7.3°$, due to strong coupling between the fundamental waveguide mode ($\alpha = 1$) and a lateral Bloch mode, which leads to a hybridization of the two modes and results in a "flattening" of the band around $\lambda \sim 460$ nm (with the shape of a "moustache") over a relatively wide angular range of $\sim 15°$, similar to what have been predicted in the literature for a plasmonic grating.[3] More details about the numerical simulations can be found in the Methods section in the main text.

**Section 2. Additional sample characterizations**

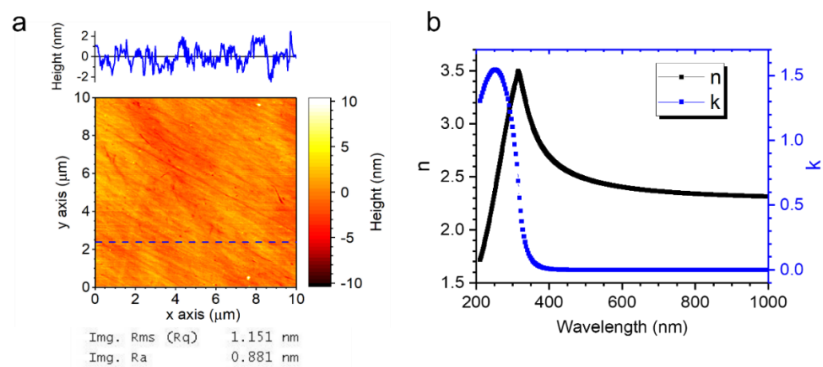

**Figure S4.** (a) AFM of a deposited $TiO_2$ film. (b) Measured optical constants ($n$ and $\kappa$) of a deposited $TiO_2$ film.

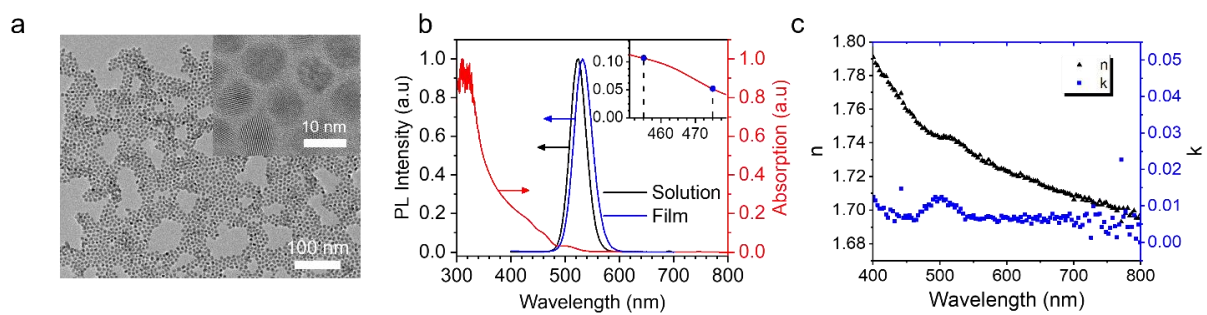

**Figure S5.** (a) TEM image of synthesised QDs. (b) Measured absorption (red) and photoluminescence of the QDs in solution (black) and in a film (blue). (c) Measured optical constants ($n$ and $\kappa$) of the QDs film.

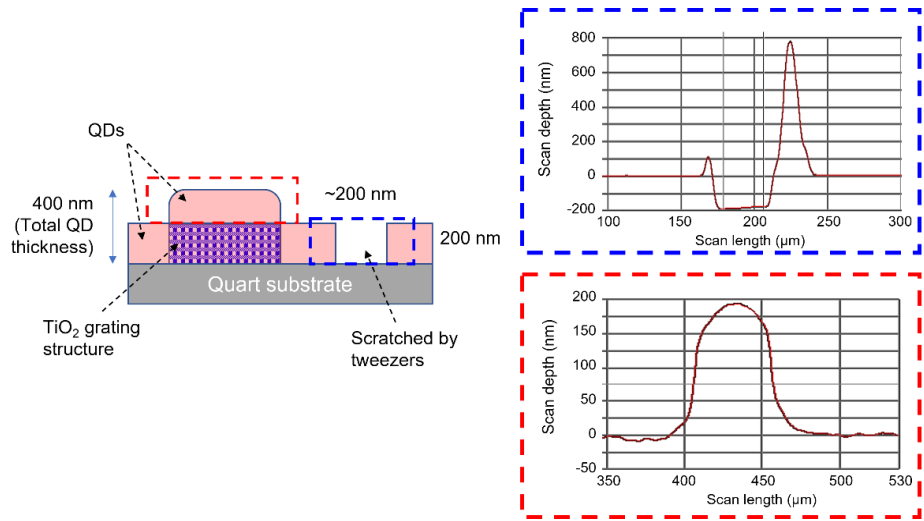

**Figure S6.** Profilometry measurement to determine the total QD film thickness in a 50×50 μm² size pixel.

**Section 3. Additional optical characterization**

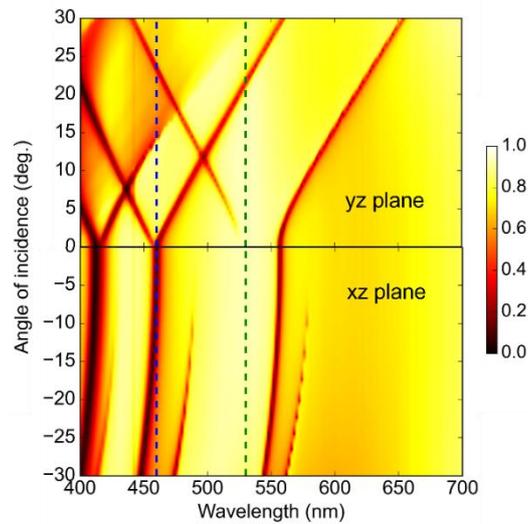

**Figure S7.** Simulated angle-resolved transmission spectrum for incident linearly polarized light along y axis. In these simulations, the dispersive values of the optical constants ($n$ and $\kappa$) of TiO$_2$ and QDs were used.

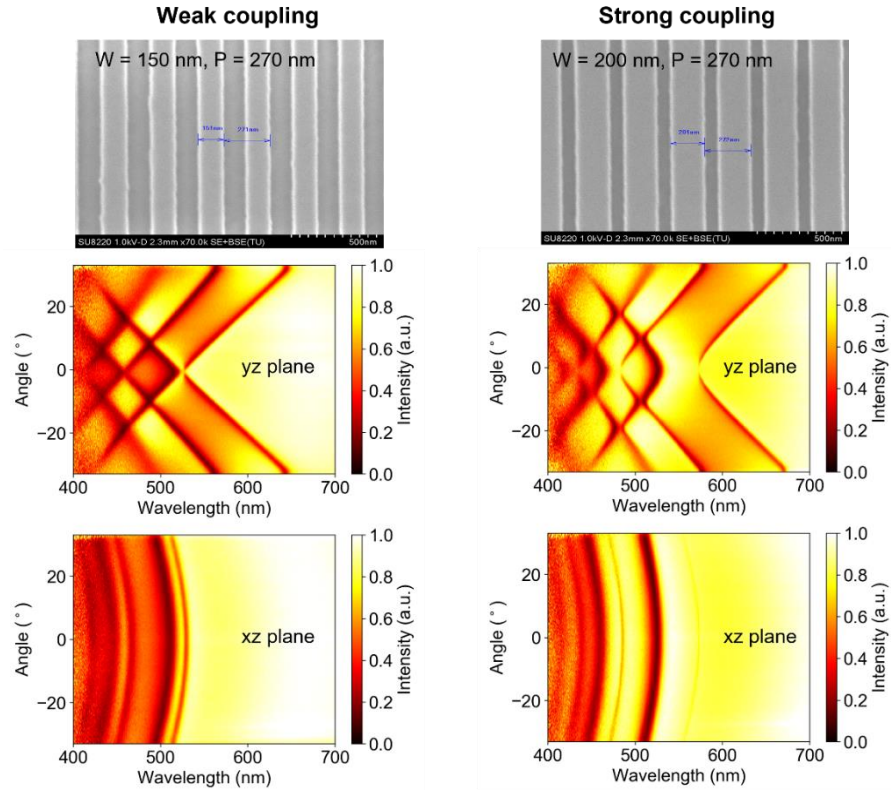

**Figure S8.** Comparison between gratings with different stripe widths (a) $W = 150$ nm (weak coupling) and (b) $W = 200$ nm (strong coupling), respectively. Top panels: SEM images of the gratings in the two cases. Middle and bottom panels: Experimental angle-resolved transmission spectra with linear polarization along the stripe direction (i.e., along x axis), in two different planes of incidence: xz plane (bottom panels) and yz planes (middle panels).

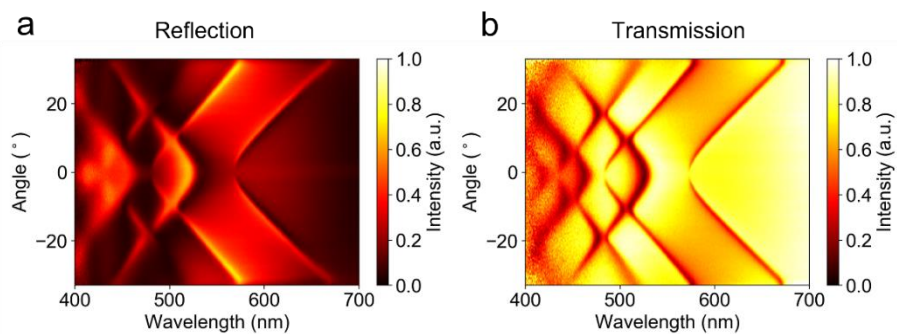

**Figure S9.** (a-b) Experimental angle-resolved reflection and transmission spectra, respectively, for the sample with same configuration as in the main text, Figure 2.

## Section 4. Absorption calculations

### A) Numerical simulations

In Figure S10a, we show the angle-resolved absorption spectrum simulated for $x$-polarized light (same configuration as in the main text, Figure 2), and its decomposition into the QDs and the TiO$_2$ absorption contributions in Figures S10b,c, respectively. As can be seen, the QDs are largely responsible for the absorption of light, which is due to a larger extinction coefficient compared to TiO$_2$ in the visible range ($\kappa \simeq 0.007$ for QDs and $\kappa \simeq 0.001$ for TiO$_2$ at $\lambda = 460$ nm) and also a better overlap with the mode electric field, as seen in Fig. 2e in the main text.

In Figure S10d, we show a "cut" at normal incidence $\theta = 0°$ of these angle-resolved spectra, which reveals that at the pumping wavelength of $\sim 460$nm, the total absorption presents a peak reaching a value of $\sim 36\%$. The total absorption also reaches a value of $\sim 67\%$ near the pumping wavelength of $\sim 460$nm at incident angle $\theta = \pm 7.3°$, as shown in Figure S10e, when the fundamental waveguide mode hybridizes with the lateral Bloch mode of the grating, leading to an even better mode overlap with the QDs (see Fig. 2e in the main text), hence increasing absorption. At these absorption peaks, the contribution by the QDs is about 93%, in both cases, the remaining 7% coming from the TiO$_2$

### B) Comparison with a bare film of QD

To appreciate what the absorption value of 67% means, obtained in our design near the pumping wavelength $\lambda \sim 460$ nm and at incident angle $\theta = 7.3°$, we can calculate the propagation distance $z_p$ necessary for the light to be attenuated (i.e. absorbed) by the same amount when propagating through a bare film of QDs. For an incident light intensity $I_0$, the intensity in a homogeneous medium decays exponentially with the propagation distance $z$ as:[5]

$I(z) = I_0 e^{-\mu z}$ where the attenuation (or absorption) coefficient $\mu$ reads $\mu = \frac{4\pi\kappa}{\lambda}$.

Therefore, to obtain an equivalent absorption of 67%, i.e. a residual light intensity of $I(z_p)/I_0 = 1 - 0.67$, at around $\lambda \sim 460$ nm where the extinction coefficient of QDs is $\kappa \simeq 0.007$, the light needs to propagate over a distance $z_p \simeq 5.8$ μm, which is to be compared to the total thickness $T = 400$ nm used in our design. In other words, thanks to a judicious nanophotonics design, we use 14.5 times less materials to achieve the same absorption level as in a bare film of QDs (in fact even less QD material are used since the QDs layer is partially filled with the TiO₂ stripes, thus reducing the amount of QDs within the layer).

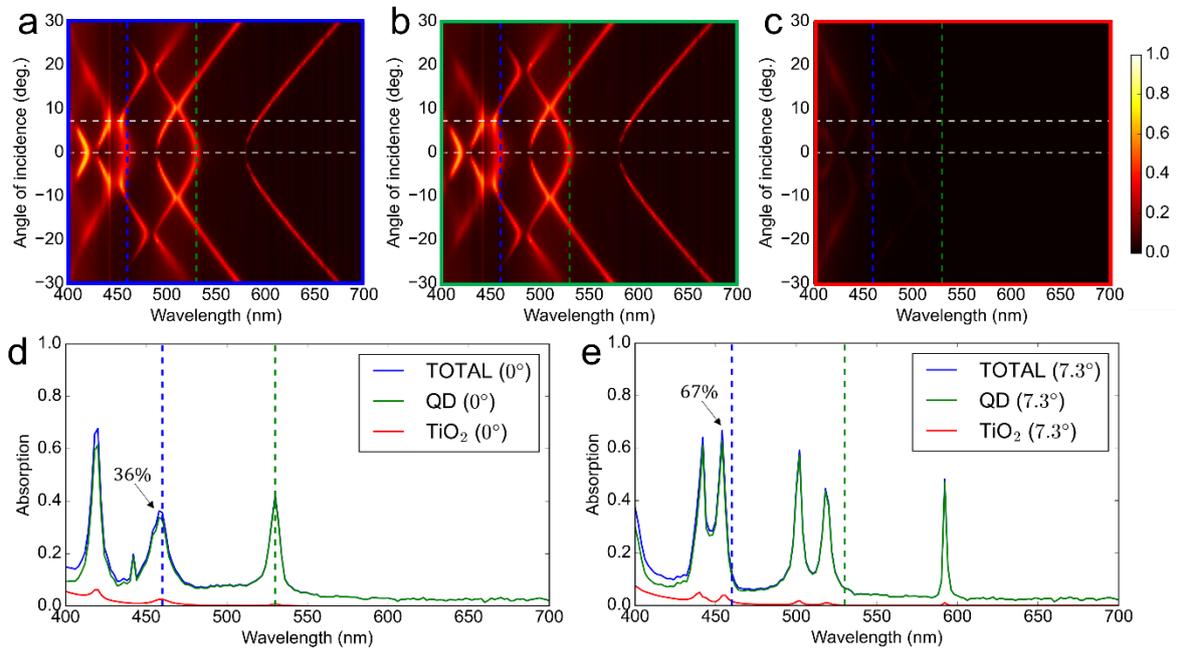

**Figure S10.** (a) Simulated angle-resolved absorption spectrum and its decomposition into (b) QDs and (c) TiO₂ contributions. The horizontal dashed white lines represent the cases of $\theta = 0°$ (lower line) and $\theta = 7.3°$ (upper line). (d-e) Total (blue line), QDs (green line) and TiO₂ (red line) absorption spectra at incidence angle $\theta = 0°$ (d) and $\theta = 7.3°$ (e). In all plots, the vertical dashed blue and green lines denote the pump and emission wavelengths $\lambda = 460$ nm and $\lambda = 530$ nm, respectively. In these simulations, the dispersive values (measured) of the optical constants ($n$ and $\kappa$) of TiO₂ and QDs were used.

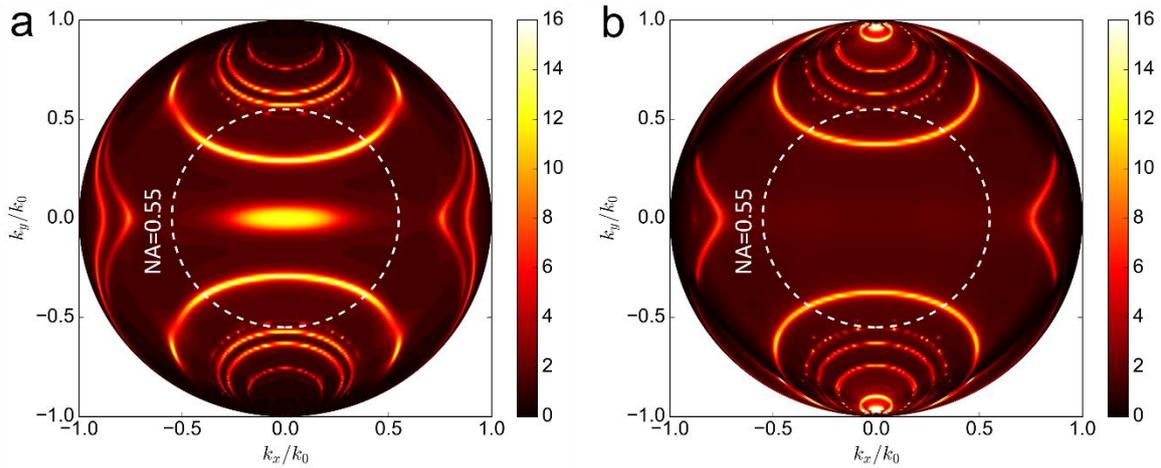

**Figure S11.** Simulated BFP emission of QDs inside the nanoantennas at $\lambda = 530$ nm, decomposed into (a) $x$-polarized and (b) $y$-polarized light emission (i.e., light polarized along and perpendicular to the stripes, respectively). The emitted power plotted here is normalized to the one emitted by a bare QD film of thickness 200 nm on glass substrate. In these simulations, the dispersive values of the optical constants ($n$ and $\kappa$) of $TiO_2$ and QDs were used.